\documentclass[10pt, conference]{IEEEtran}
\IEEEoverridecommandlockouts
\usepackage{cite}
\usepackage{amsmath,amssymb,amsfonts}
\usepackage{enumitem}

\usepackage{algorithm}
\usepackage[noend]{algpseudocode}

\usepackage{graphicx}
\graphicspath{ {./images/} }
\usepackage{textcomp}
\usepackage{xcolor}
\def\BibTeX{{\rm B\kern-.05em{\sc i\kern-.025em b}\kern-.08em
    T\kern-.1667em\lower.7ex\hbox{E}\kern-.125emX}}
\begin{document}

\title{Reliability Model for Incentive-Driven IoT Energy Services
}

\author{\IEEEauthorblockN{Amani Abusafia}
\IEEEauthorblockA{\textit{School of Computer Science} \\
\textit{The University of Sydney}\\
Sydney, Australia \\
amani.abusafia@sydney.edu.au}
\and
\IEEEauthorblockN{Athman Bouguettaya}
\IEEEauthorblockA{\textit{School of Computer Science} \\
\textit{The University of Sydney}\\
Sydney, Australia \\
athman.bouguettaya@sydney.edu.au}
}

\maketitle

\begin{abstract}
 We propose a novel \textit{reliability} model for composing energy service requests. The proposed model is based on consumers' behavior and history of energy requests. The reliability model ensures the maximum incentives to providers. Incentives are used as a green solution to increase IoT users' participation in a crowdsourced energy sharing environment. Additionally, adaptive and priority scheduling compositions are proposed to compose the most reliable energy requests while maximizing providers' incentives. A set of experiments is conducted to evaluate the proposed approaches. Experimental results prove the efficiency of the proposed approaches. 

\end{abstract}


\section{Introduction}

\emph{Internet of Things (IoT)} is a paradigm of connecting everyday objects, known as \textit{things}, to the Internet. The number of connected IoT is expected to increase by 12\% every year to reach 125 billion in 2030 \cite{markit2017internet}. The ``things'' include smartphones, wearables, and vehicles. Things usually have augmented capabilities including sensing, networking, and processing \cite{whitmore2015internet}. The pervasiveness of IoT provides opportunities to crowdsource their capabilities \cite{bouguettaya2017service}.

\emph{Crowdsourcing} IoT devices integrate things to create applications by utilizing their data and functionality \cite{ziegler2015internet}. Examples of these applications include environmental monitoring , smart homes and smart cities \cite{safia2019optimising}. Abstracting crowdsourced IoT using the service paradigm may provide novel IoT services \cite{bouguettaya2017service}. \emph{IoT services} are defined by their \emph{functional} and \emph{non-functional} properties. Example of \emph{IoT services} includes WiFi sharing and energy sharing \cite{lakhdari2020Elastic}. Of particular interest is the use of energy sharing services.


Energy sharing service, also known as \emph{Energy-as-a-Service (EaaS)}, refers to the exchange of wireless energy among IoT devices \cite{lakhdari2018crowdsourcing}. An \emph{energy provider} refers to a $thing$ that can share energy\footnote{We will use ``owner'' and ``provider'' to refer to the IoT device's owner}. An \emph{energy consumer} is a $thing$ that needs energy.  Wearables such as smart textile or smart shoes may \textit{harvest} energy from natural resources, e.g. body heat or kinetic activity \cite{choi2017wearable} \cite{gorlatova2015movers}. For instance, the PowerWalk kinetic energy harvester produces 10-12 watts on-the-move power \cite{PowerWalk}. A user wearing a PowerWalk harvester can generate enough energy to charge up to four smartphones from an hour walk at a convenient pace. The harvested energy can be shared with nearby IoT devices as EaaS. EaaS may be achieved with the recently developed ``Over-the-Air wireless charging'' technology \cite{OvertheAirCharger}. For example, ``Energous'' developed a technology to enable wireless charging up to a distance of 15 feet\footnote{https://www.energous.com/}.  In this paper, we focus on using wearables and IoT devices as energy providers. 

Providing energy services (EaaS) has several advantages: EaaS is a crowdsourced \emph{green} solution as it reduces carbon footprint by utilizing spare and renewable energy \cite{lakhdari2018crowdsourcing}. EaaS, moreover, offers spatial freedom and convenience as an alternative to carrying power banks or plugging into a power outlet. Moreover, there are multiple domains where EaaS is necessary due to the lack of power outlets. These domains include Disaster control, entertainment, and crisis response \cite{PowerWalk}. Additionally, the concept of \textit{wireless crowdcharging} has received increasing interests recently \cite{bulut2018crowdcharging}\cite{lakhdari2020fluid}. The wireless charging market is expected to reach  \$5B in 2022 \cite{NoofConnectedDevices}. This recent trend towards the adoption of wireless charging technology and the ubiquity of IoT will result in the emergence of EaaS.

 The prospective EaaS environment consists of confined areas also known as microcells. A microcell may be defined as any area where people gather such as coffee shops,  movie theaters, and restaurants. Things, in this environment, may share energy using the Energy-as-a-Service model. \emph{Energy-as-a-service model}, as previously stated, is represented by its \emph{functional} and \emph{non-functional}, Quality of service (QoS), properties \cite{lakhdari2020composing}. The function of the energy service is is the wireless delivery of  energy among IoT devices. The non-functional (QoS) properties include the amount, location, and duration of energy. To the best of our knowledge, existing research mainly focuses on the composition of EaaS from a consumer's aspect. {We focus on the selection and composition of energy service requests from a provider's prospective}. The existing research in composing requests has a different context than the EaaS environment including the mobility  of IoT users and the resources constrains \cite{abusafia2020incentive}.

The prospective EaaS environment consists of confined areas, also known as microcells. A microcell may be defined as any area where people gather, such as coffee shops,  movie theaters, and restaurants. Things, in this environment, may share energy using the Energy-as-a-Service model. \emph{Energy-as-a-service model}, as previously stated, is represented by its \emph{functional} and \emph{non-functional}, Quality of service (QoS), properties \cite{lakhdari2020composing}. The function of the energy service is the wireless delivery of energy among IoT devices. The non-functional (QoS) properties include the amount, location, and duration of energy. To the best of our knowledge, existing research mainly focuses on the composition of EaaS from a consumer's aspect. {We focus on the selection and composition of energy service requests from a provider's prospective}. The existing research in composing requests has a different context than the EaaS environment, including the mobility of IoT users, and the resources constraints \cite{abusafia2020incentive}.

IoT environments are highly dynamic, i.e.,  IoT devices move freely between microcells. Movements in and out of microcells are categorized as either \textit{voluntary} or \textit{involuntary}, which would result in disrupting the energy transfer. Voluntary disruption is an \textit{explicit} decision by the consumer to discontinue the energy transfer.  Involuntary disruption refers to an \textit{unintended} temporary discontinuity in the energy transfer by the consumer. In the case of voluntary disruption, a selected energy request may be discontinued for several reasons, including cancellation of the request or leaving the range of energy transfer.  In the case of involuntary disruption, a consumer may unintentionally leave intermittently, e.g., moving in and out the microcell. The disruption of a selected request may reduce the amount of shared energy, and therefore, affect the provider's reward. Therefore, a re-composition is required to ensure that the provider is still getting an optimal reward. However, there is no guarantee that this would be achieved because an optimal replacement is dependent on the availability of other consumers' requests.

Reliability is proposed as a measure to gauge the level of commitment from energy consumers. Commitment is defined as the {\em  likelihood} of a consumer completing the consumption of an energy request. We will use past consumers' history to assess the level of commitment. We propose a {\em consumer-side} reliability model using the level of consumers' commitment as a key indicator for computing reliability. \emph{We focus on selecting and composing the best set of energy requests using our reliability model}. 

The main contributions of this paper are:
\begin{itemize}
    \item A novel reliability model considering consumers' history and mobility model. 
    \item A spatio-temporal reliable selection of energy requests.
    \item A scheduling reliability-based incentive-driven composition of energy service requests.
    \item An adaptive composition to maximize the reliability and the reward of the provider.
\end{itemize}

\begin{figure}[!t]
\centering
\includegraphics[width=\linewidth]{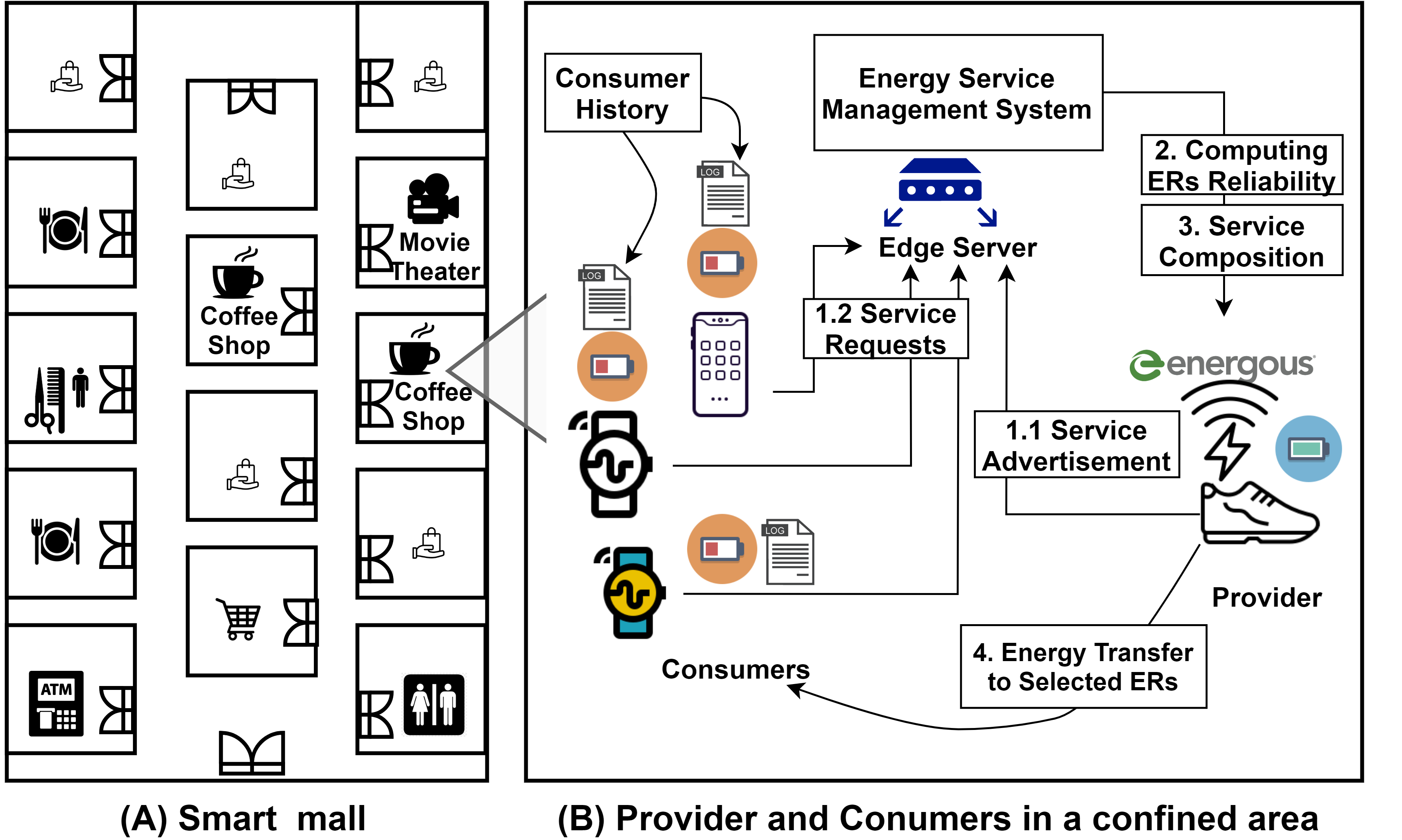}
\caption{Reliability-Based Wireless Energy Sharing System}
\label{Scenariofig}
\end{figure}

\textbf{{Motivating Scenario:}} We consider the scenario as shown in Fig.\ref{Scenariofig}. Assume that a Mall is split into geographical $microcells$ such as cafes, restaurants, movie theaters, etc. (see Fig.\ref{Scenariofig}A). We also assume there are several IoT devices and wearables (provider and consumers) in a microcell (see Fig.\ref{Scenariofig}B). The devices are assumed to be equipped with wireless energy transmitters and receivers such as ``Energous''\footnote{https://www.energous.com/}. All local energy requests and advertisements are submitted and handled at the edge, e.g., a router associated with the microcell. We assume the system puts a limit on the amount of requested energy by a consumer. We also assume that the IoT coordinator uses a reward system to encourage providers to share energy. Rewards come in the form of stored credits to providers. The collected credits may be used later by the provider to increase the limit on the amount of requested energy when they are in the consumer role. We assume that each provider may also be a consumer and vice-versa. A provider receives a reward based on the amount of shared energy. Assume a \textit{device owner}, e.g., a \textit{smart shoe}, wishes to share their spare energy with other IoT devices in the microcell. In this scenario, the provider is assumed to be able to receive requests from multiple energy consumers via the edge. We assume that the provider can transfer energy up to one consumer at a time. A service consumer may not commit to receiving all the requested energy. In this respect, the reliability of such consumers is paramount: Unreliable consumers may hinder the provider from utilizing their spare energy and thus reduce their reward. The unreliability of consumers might be voluntary, e.g., intentionally canceling the request or involuntary, e.g., moving in and out of the microcell. 

\begin{figure}[!t]
\centering
\includegraphics[width=1\linewidth]{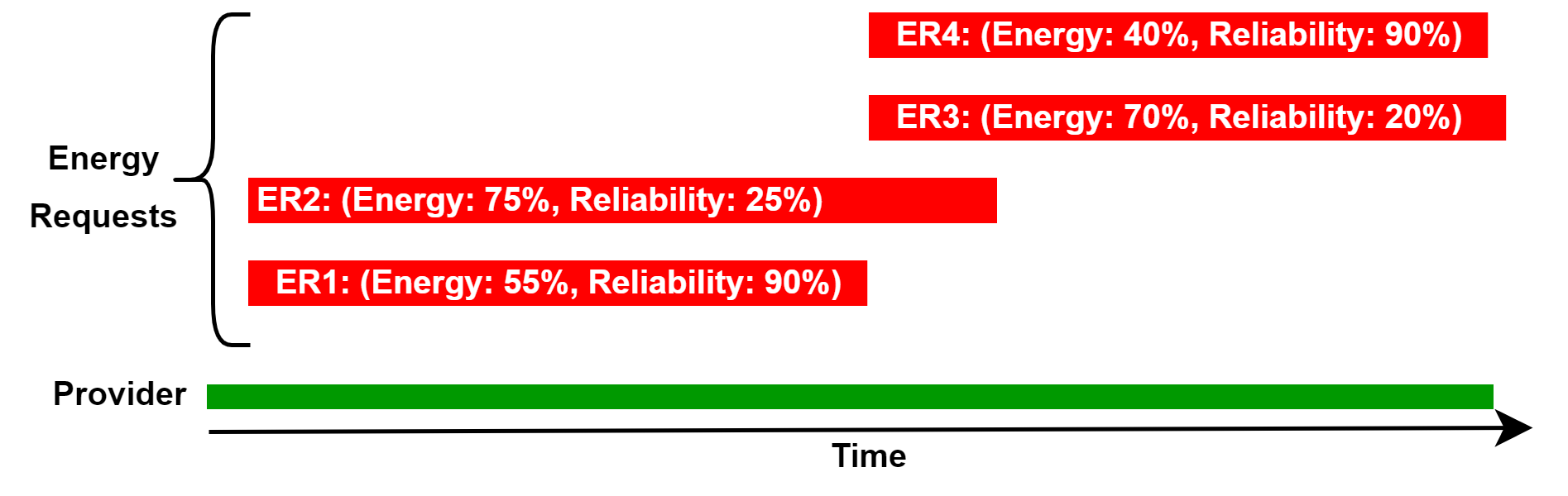}
\caption{Example of Energy Requests with reliability}
\label{EnergyRequests}
\end{figure}

Fig.\ref{EnergyRequests} represents an example of received energy requests with different reliability. The first and second energy requests ($ER1$ and $ER2$) require 55\% and 75\% of the provider energy, respectively. $ER1$ has a 90\% reliability. $ER2$ has a 25\% reliability. Selecting $ER2$ will give a bigger reward for the provider than $ER1$ but, $ER2$ is more likely to be canceled.  If the request gets canceled, the reward of the provider may be reduced. {It is\textit{ challenging} to replace a canceled request as there is no guarantee to find another available request}. Evaluating the reliability of the energy requests before selection may enhance the chances of obtaining the most committed $ERs$, and therefore, offer the best reward for the provider.

We focus on {composing reliable energy service requests to maximize the reward of the energy provider}. We assume a single energy provider may provide a \emph{specific} time interval, during which they may share their energy with multiple energy consumers. The \emph{composition} of energy service requests includes selecting the most reliable requests that maximize the provider's reward. The reliability of energy requests is calculated using a reliability model. Note that we assume a provider is static during the composition.  Additionally, we assume consumers might move during the composition. To the best of our knowledge, no algorithm addresses the problem of energy requests' reliability in composing energy service requests to incentivize the provider.

\section{System Model and Definitions}
\label{SysModel}
We propose a formal model of our reliable composition of energy requests. {We consider the scenario of energy sharing in a microcell during a particular time interval} \emph{T}. We use the below definitions to formulate the problem. 

\textbf{Definition 1:  Energy-as-a-Service (EaaS).} We adopt the definition of EaaS in \cite{lakhdari2018crowdsourcing}.  An EaaS is defined as a set of $\{E\_{id}, E\_{Pid}, F,Q\}$, where:
\begin{itemize}
    \item $E\_id$ is an energy service's unique identifier
    \item $E\_{Pid}$ is a unique provider's identifier
    \item $F$ is the function of sharing energy by an IoT device
    \item $Q$ is the non-functional , Quality of Service ($QoS$), attributes including:
        \begin{itemize}
            \item $P\_{ec}$ is the capacity of energy the provider may share
            \item $P\_{loc}$ is the  location of the provider $<x,y>$
            \item $P\_{st}$ is the start time of the provider's stay 
            \item $P\_{et}$ is the end time of the provider's stay 
        \end{itemize}
\end{itemize}

\textbf{Definition 2: Energy Service Request (ER).} We extend the existing definition of ER model \cite{abusafia2020incentive} by considering the reliability of the ER. An ER request is defined as a set of $\{ER\_{id}, ER\_{Cid}, F,Q\}$, where:
\begin{itemize}
    \item $ER\_id$ is an energy request's unique identifier
    \item $E\_{Cid}$ is a unique consumer's identifier
    \item $F$ is the function of receiving energy by an IoT device
    \item $Q$ is the non functional $QoS$ attributes including:
    \begin{itemize}
        \item $C\_{re}$ is the amount of requested energy by the consumer
        \item $C\_{st}$ is the start time of the consumer's stay
        \item $C\_{et}$ is the end time of the consumer's stay
        \item $C\_{loc}$ is the  location of the consumer$<x,y>$
        \item $C\_{Rel}$ is the reliability score of the consumer
\end{itemize}
\end{itemize}

\textbf{Definition 3: Incentive.} The energy service provider gains an incentive reward $R$ by sharing energy. Rewards are provided by {the IoT coordinator at the edge}. The rewards come in the form of stored credits to providers.  We assume the system has a  limit for the amount of requested energy. The collected credits may be used later by the provider to increase the limit of requested energy as a consumer. There has been a study on the use of incentives to increase energy provision participation \cite{abusafia2020incentive}. The study shows that the amount of provided energy affects the value of the incentive to provide energy. As a result,  we will compute the reward ($rwd$) of an energy request based on the amount of requested energy using the below equation.

\begin{equation}
 \label{rewardEquation}
 rwd  = C\_{re} / P\_{ec} 
\end{equation}
where $ C\_{re} $ is the amount of requested energy and $P\_{ec}$ is the capacity of energy the provider may share. 

\textbf{Definition 4: Reliability-based energy service requests composition problem.}
We assume in a microcell, there exists an energy service $EaaS$ and a set of $n$ energy requests $ERs = \{ER_{1},ER_{2},....,$ $ER_{n}\}$ as shown in Fig.\ref{EnergyRequests}. The $EaaS$ is posted by a provider $P$. $ER$ are initiated by consumers $C$. $EaaS$ and $ER$ are described using Definitions 1 and 2, respectively. {We reformulate the energy sharing problem as a service composition problem using the service model.} The composition of energy requests considers the spatio-temporal features of the service and requests. {Composing energy requests for a provider's} $EaaS$ requires the composition of energy requests $ER_{i} \in ER$ where $[C\_{st_{i}},C\_{et_{i}}] \subset [P\_{st},P\_{et}]$, $\sum C\_{re_{i}} \geq P\_{ec}$, the provider reward = $\sum rwd_{j}$ is the maximum they can get and the reliability of the composition = $\sum C\_{Rel_{i}}$ is the highest in reliability.

\noindent We use the following assumptions to formulate the problem:

\begin{itemize}[ noitemsep,nosep,leftmargin=20pt,labelsep=8pt,itemindent=3pt, labelwidth=20pt]
\item The IoT devices are equipped with wireless energy transmitters and receivers.
\item The composition considers the case of a \textit{single} provider and \textit{multiple} consumers.
\item The provider has fixed energy during the composition.
\item The provider and consumers may have different time windows, but consumers' time window must fall within the provider time window $T_c \in T_p$.
\item The provider has fixed location $<x,y>$ for the duration of the energy service.
\item The provider transfers energy within their range (15 feet using Energous) to one consumer at a time.
\item There is no energy waste while sharing. 
\item All consumers' have a history of energy requests 
\item Consumers' requests include information about the reward they are willing to pay. 
\item A trust framework has been implemented to preserve the privacy of the participating IoT devices.

\end{itemize}
\section{Reliability Model}
\label{RM}
We define the reliability model to determine the reliability score ($Rel$) of the energy requests.  The reliability model considers the behavior of consumers and their history in requesting energy. We assume that the history of consumers will be collected in a phase prior to deploying our proposed framework. The below attributes are used to compute the reliability score.
\begin{itemize}[leftmargin=8pt]
\item \textbf{Consumption  History:} Consumers history of energy requests may be an indication of their reliability. The number of successful requests can be used in computing the reliability. A successful request is one that received all that was requested.  Therefore, the computation of a consumer's  reliability based on their history ($Rel\_{History}$) is:
\begin{equation}
Rel\_{History} = SER/Total\_ER 
\end{equation}
 Where $SER$ is the consumer's number of successful requests and $Total\_ER$ is the consumer's total number of submitted requests.

\item \textbf{Energy Disruption}: 
As previously mentioned, disruption of energy transfer may occur voluntary or involuntary. We refer to the disconnection in energy transfer without reconnecting as the voluntary disruption (See Fig.\ref{voluntary}). Similarly, we refer to the intermittent disconnection in energy transfer within one energy request as the involuntary disruption (See Fig.\ref{involuntary}). We define the reliability computation of both types of energy disruption as follows:
\begin{itemize}[leftmargin=0pt]
\item \textbf{Voluntary Reliability:}\label{voluntaryDef}
Voluntary disruption refers to a conscious decision by the consumer to discontinue the energy transfer.  We identify voluntary disruption to have a  disconnection in the energy request without reconnecting (See Fig.\ref{voluntary}). A disconnection's duration has to be greater than a predefined threshold ($\delta$) to be considered as a voluntary disruption. In contrast, the disconnection is not considered as a voluntary disruption if the duration of the disconnection is less than a predefined threshold ($\epsilon$). If a voluntary disruption is identified in a previous energy request, we compute its voluntary reliability. We measure the voluntary reliability ($Rel\_{Voluntary}$) based on energy and time. The reliability of energy ($Rel\_{Energy}$) for a given consumer is:
\begin{equation}
Rel\_{Energy} =   \left(\sum_{i=1}^{n} \left( \frac{ Received\_Energy_{i}}{Requested\_Energy_{i}} \right) \right) /n 
\end{equation}
where \textit{n} is the number of the previous ER for a given consumer. The previous equation measures the average energy ratio of all the consumer's previous ER.
 \begin{figure}[!t]
\centering
\includegraphics[width=\linewidth]{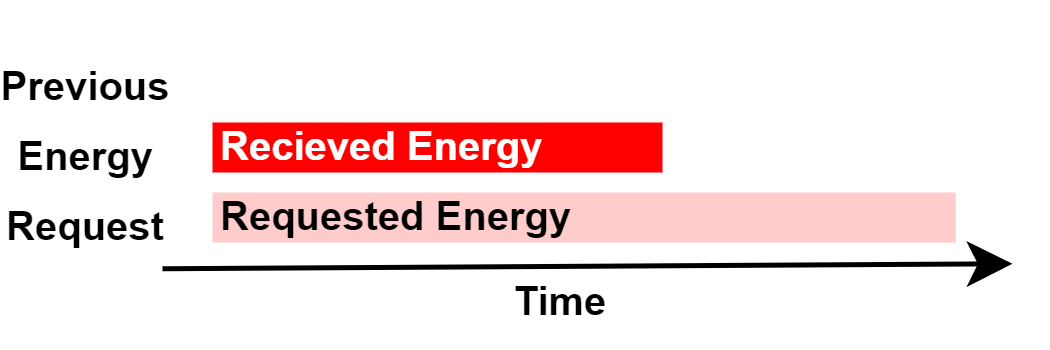}
\caption{Example of voluntary disruption. }
\label{voluntary}
\end{figure}

The reliability of time ($Rel\_{Time}$) for a given consumer is:
\begin{equation}
Rel\_{Time} =   \left(\sum_{i=1}^{n} \left( \frac{Stay\_Time_{i}}{Required\_Time_{i}} \right)\right)/n \end{equation}
where \textit{n} is the number of the previous ER for a given consumer. The previous equation measures the average time ratio of all the consumer's previous ER. 

The equation of voluntary reliability ($Rel\_{Voluntary}$) is:
\begin{equation}
\label{voluntrayeq}
Rel\_{Voluntary} = (Rel\_{Energy} + Rel\_{Time} ) /2
\end{equation}

\item \textbf{Involuntary Reliability: }
Involuntary disruption refers to an intermittent disconnection in energy transfer within a single ER (See Fig.\ref{involuntary}). We identify involuntary disruption to have at least one disconnection between two connections of the same ER. A disconnection's duration has to be greater than a predefined threshold ($\delta$) to be considered as an involuntary disruption. In contrast, the disconnection is not considered as an involuntary disruption if the duration of disconnection less than a predefined threshold ($\epsilon$). Moreover, we consider a single disconnection at the end of the energy transfer to be an involuntary disruption if its duration is less than ($\alpha$). 
If an involuntary disruption is identified in a previous ER, we compute its involuntary reliability. The involuntary reliability for a ER ($Rel\_{Involuntary}$) is computed based on the frequency of disconnections and the reliability of the duration of disconnections in the request. ER is more reliable when the disconnections are less frequent and for smaller time periods. For every previous ER ($ Prev\_ER$), we compute the reliability based on the frequency of disconnections ($Rel\_f$) as:
\begin{equation}
\resizebox{\hsize}{!}{$
 Rel\_f_i =
\begin{cases}
    1 & \text{if $f_{i} = 0$} \\
    0.7, & \text{if }
    
    \!\begin{aligned}[t]
      (f_{i}&= 1 \& \delta < d_{i} < \epsilon \&
      d_{i}\_{et} \neq Prev\_ER_{i}{\_}_{et} ) or \\
      (f_{i}&= 1 \&   d_{i} < \alpha \& 
      d_{i}\_{et} = Prev\_ER_{i}{\_}_{et})
      \end{aligned}
    \\
    \frac{1}{f_{i}} & \text{if} f_{i} >= 1 \& \delta < d_{i} < \epsilon  \\
\end{cases}$}
\end{equation}

where $f_{i}$ is the frequency of disconnections within a single previous ER ($ Prev\_ER_i$), $d_{i}$ is the duration of disconnections within single ER, and $Prev\_ER_{i}{\_}_{et}$ is the end time of the ER. 

\begin{figure}[!t]
\centering
  \includegraphics[width=0.8\linewidth]{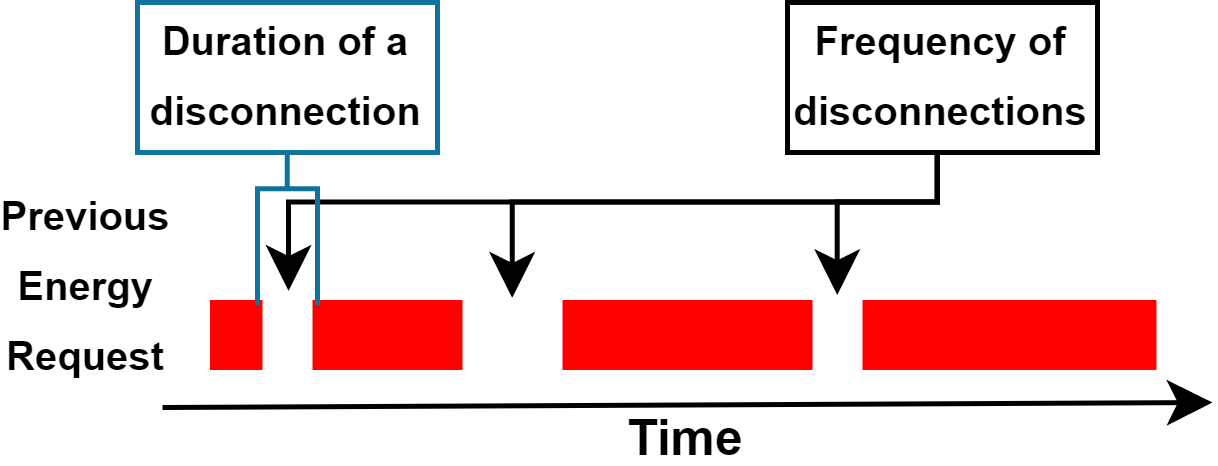}
\caption{Example of involuntary disruption. }
\label{involuntary}
\end{figure}

We compute the reliability of a consumer,  based on the frequency of disconnections, as the average of the reliability of frequency ($Rel\_f$) of  all their previous requests ($Prev\_ER$). Therefore, he equation to compute the reliability of a consumer based on the frequency of disconnections ($Rel\_{freq}$) is: 
\begin{equation}
  Rel\_{freq} =  \left( \sum_{i=1}^{n} Rel\_{f}_{i}\right)/n
\end{equation}
where \textit{n} is the number of the previous requests for a given consumer. 

The duration of disconnections in ER may results in a loss in the provider's reward. For every previous ER, we compute the reliability based on the duration of disconnections  ($Rel\_{d}$) as:
\begin{equation}
 Rel\_{d}_{i} = 1 - \frac{\sum_{j=1}^{k} {d_{{i}_{j}}}}{Prev\_ER_{i}{\_}_{et} - Prev\_ER_{i}{\_}_{st}}
\end{equation}

where $k$ is the number of disconnections in a consumer's previous ER,  $d_{{i}_{j}}$ is the duration of disconnection within a single ER, $Prev\_ER_{i}{\_}_{st}$ and $Prev\_ER_{i}{\_}_{et}$ are the start and end time of the ER, respectively. 

We define the reliability of the consumer ,  based on the duration of disconnections, as the average of the reliability of duration of all their previous requests ($Prev\_ER$). Therefore, the equation to compute  the reliability of a consumer based on the duration of disconnections ($Rel\_{dur}$) is the following:
\begin{equation}
  Rel\_{dur} =  \left(\sum_{i=1}^{n} Rel\_{d}_{i}\right)/n
\end{equation}
where \textit{n} is the number of the previous ER for a given consumer. 

As previously stated, the involuntary reliability of a consumer will be computed based on the reliability of the frequency ($Rel\_{freq}$) and the reliability of the duration of disconnections ($Rel\_{dur}$). Therefore, we compute the involuntary reliability of a consumer ($Rel_{Involuntary}$) as the following:

\begin{equation}
Rel\_{Involuntary} =  (Rel\_{freq} + Rel\_{dur})/2
\end{equation}
\end{itemize}

\begin{figure}[!t]
\centering
\includegraphics[width=0.9\linewidth]{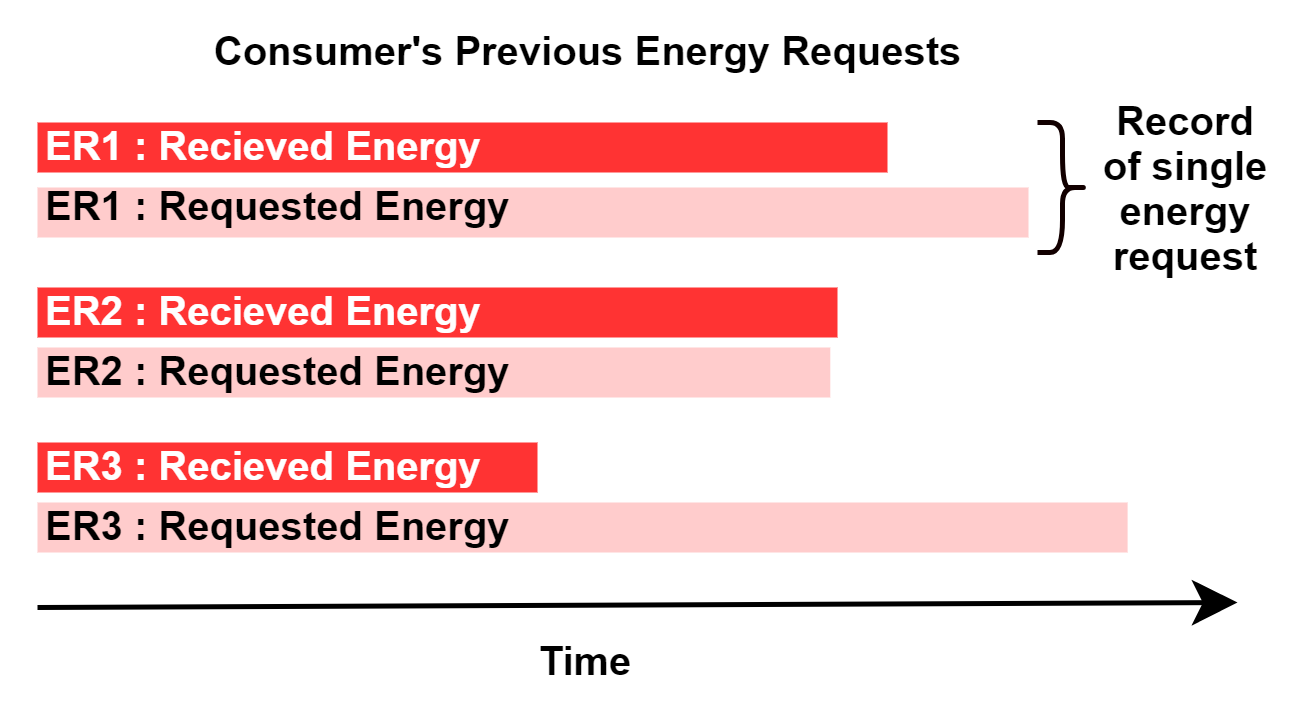}
\caption{Example of random behavior}
\label{RandomBehavior}
\end{figure}

\item \textbf {Random Behavior:}
Random behavior measures the level of fluctuations regarding the consumer's commitment to receive the requested energy (See Fig.\ref{RandomBehavior}). We use the variation rate to estimate the random behavior of a consumer.  Consumers with a high variation rate of commitment will be identified with random behavior. A highly random behavior is an indication of an unreliable consumer. We measure the variation rate of a consumer's behavior based on energy and time. The equation of the Energy variation rate ($E\_vr$) for a given consumer is:
\begin{equation}
\label{ERS}
E\_vr =  \sqrt{ \frac{ \sum_{i=1}^{n}\left( \frac{Received\_Energy_{i}}{Requested\_Energy_{i}} - \overline{\left(\frac{Received\_Energy}{Requested\_Energy}\right)}\right)^2}{n} }
\end{equation}

where \textit{n} is the number of the previous ER for a given consumer. The previous equation measures, for a given consumer, the variation between the energy ratio of their previous ER and the mean of all their previous ER.

The equation of the Time variation rate ($T\_vr$) for a given consumer is:
\begin{equation}
\label{TRS}
T\_vr =  \sqrt{ \frac{ \sum_{i=1}^{n}\left( \frac{Stay\_Time_{i}}{Required\_Time_{i}} - \overline{\left(\frac{Stay\_Time}{Required\_Time}\right)}\right)^2}{n} }
\end{equation}

where \textit{n} is the number of the previous ER for a given consumer. The previous equation measures, for a given consumer, the variation  between the time ratio of their previous ER and the mean of all their previous ER.

As previously mentioned, the random behavior of a consumer is measured based on energy and time. Therefore, we compute the reliability of a consumer based on their random behavior ($Rel\_{Random}$) as:
\begin{equation}
Rel\_{Random} = (2 -(E\_vr + T\_vr) )/2
\end{equation}
{ Where $E\_vr$ is the energy variation rate computed by Eq.\ref{ERS} and $T\_vr$ is the time variation rate computed by Eq.\ref{TRS}.}
\end{itemize}
\textbf{Consumer Reliability:} The $reliability$ score  of a consumer is computed as the summation of all aforementioned reliability attributes. Therefore, the equation of computing the reliability score of a consumer ($C_{Rel}$) is:
\begin{equation}
\label{consumerReleq}
  C_{Rel} = \sum_{i \in n}^{}w_{i}\times Rel\_{i} 
\end{equation}
 Where $n =\{History, Voluntary, Involuntary, Random\}$ and $w$ is the weight of each reliability attribute. We use  $w$ to normalize the impact of each factor on the overall reliability score. 
 
\noindent\textbf{Service Composition Reliability Score:}
The reliability score of a service composition will be the summation of the consumers' reliability scores of the selected requests. Therefore, the equation of computing the reliability score of a service composition is:
\begin{equation}
\label{compReliabilityEq}
Service \;Composition\; Reliability = \sum_{j=1}^{m} {C_{Rel}}_{j} 
\end{equation}
Where $m$ is the number of selected energy requests that would receive the energy from the provider, and ${C_{Rel}}_{j}$ is the reliability score of the consumer of a selected energy request.

\noindent\textbf{Actual Reward:} The actual reward of an ER refers to the reward of a request considering its consumer's reliability score. For example, if an ER's consumer reliability score is 80\%, then this indicates that the consumer will probably commit to 80\%  of the requested energy. Therefore, we compute the actual reward for an ER ($Act\_{Rwd}$) with respect to its consumer's reliability as the following:
\begin{equation}
\label{ActRwd}
Act\_{Rwd} = rwd * C_{Rel}
\end{equation}

Where $rwd$ is the reward of the ER computed using Eq.\ref{rewardEquation}  and $C_{Rel}$ is the reliability score of the consumer of an ER computed using Eq.\ref{consumerReleq}.

\begin{figure}[!t]
\centering
\includegraphics[width=\linewidth]{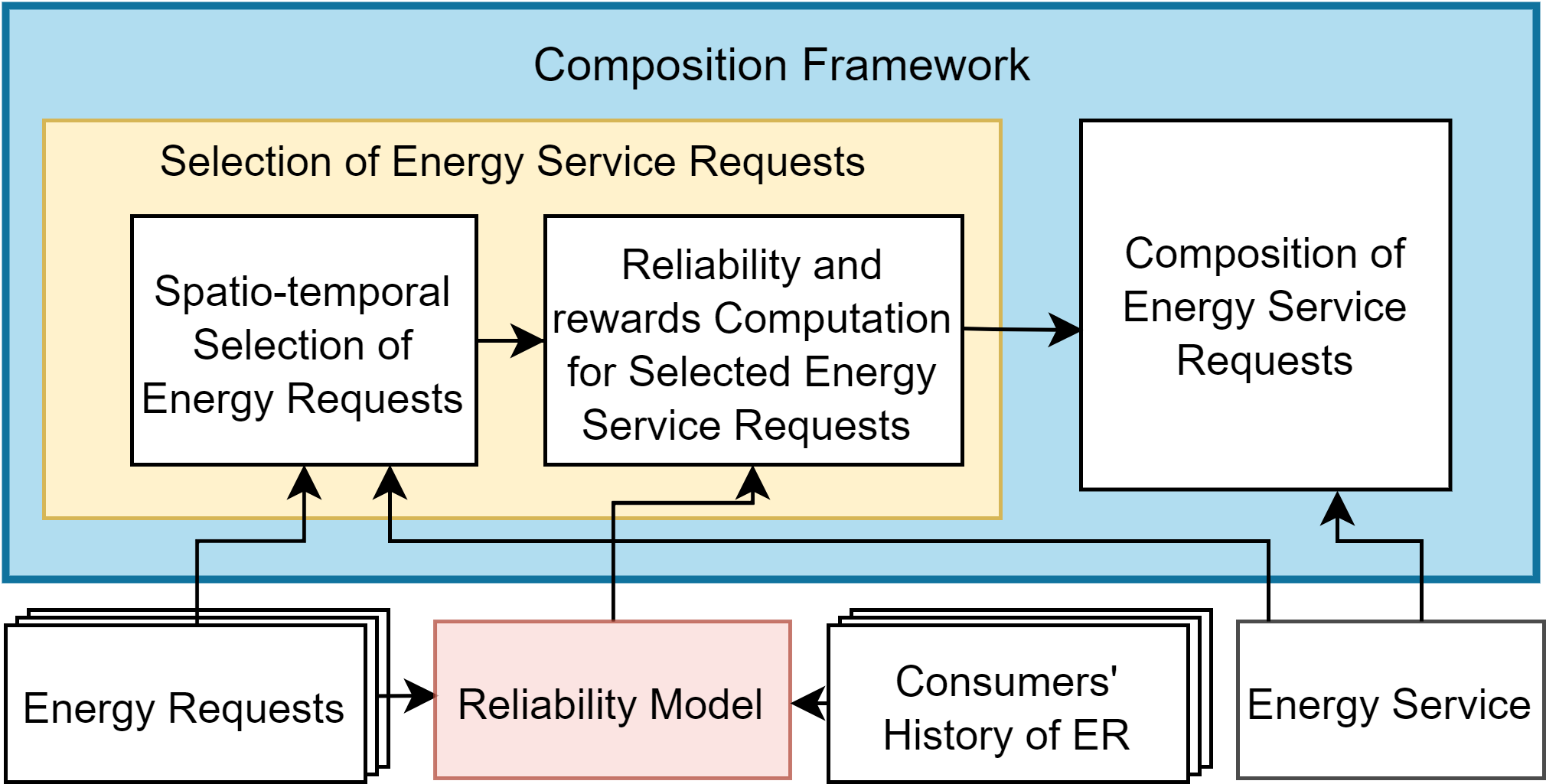}
\caption{The composition framework for ER services.  }
\label{ReliableEnergyFrameWork}
\end{figure}

\section{Composition Framework for Energy Requests}

 The framework of composing energy service requests will involve two phases: Selecting energy service requests and composing energy service requests (See Fig.\ref{ReliableEnergyFrameWork}). The first phase includes the selection of the energy requests (ERs) that can be composed, within space and time, with the energy service (EaaS). Then, we calculate the reliability score for each of the selected ERs using the reliability model.  Additionally, the reward of each ER is computed in the same phase using Eq.\ref{ActRwd}. The second phase is the composition of the most reliable ERs that maximize the provider's reward. Our composition is designed for the case of a \textit{single} provider and \textit{multiple} consumers. The composition is inspired by Priority-based scheduling algorithms \cite{kruse2007data}.
 
 \begin{algorithm}[t!]
     \renewcommand{\algorithmicrequire}{\textbf{Input:}}
    \renewcommand{\algorithmicensure}{\textbf{Output:}}
    \caption{Reliability-Based Composition of ER Services}
    \label{alg:RCER}
    \begin{algorithmic}[1]
        \Require $EaaS, All_{ER}, Reliability\; Model (RM) $
        \Ensure $ER_{comp} , compRel , compRwd$ 
        \Statex \textbf{Phase 1: Selection of Energy Service Requests}
        \For{$ER_{i}\; in\; All_{ER}$}\label{startph1} \do 
            \State \If{$ [C_{st},C_{et}] \subseteq [P_{st}, P_{et}]$}
          
                \small{$ d = Compute\_Distance(C_{loc.x},C_{loc.y},P_{loc.x},P_{loc.y}) $}
                         
                \If{$d \leq Max\;Energy\;Distance$ }
                    \If{$C_{i.re}\leq P_{ec}$}
                        \State $C_{i.Rel} = Compute\_Reliability(ER_{i},RM)$     
                        \State $ER_{i.Rwd} = C_{i\_re} /P\_{ec}$
                        \State $ER_{i.Act\_Rwd} = C_{i.Rel} * ER_{i.Rwd}$
                        \State \small{$SelectedERs.insert(ER_{i} , C_{i.re},ER_{i.Act\_Rwd})$}\label{endph1}
                    \EndIf 
                \EndIf 
            \EndIf
        \EndFor
        \Statex \textbf{Phase 2: Composition of Energy Service Requests}
        \State $SelectedERs_{sorted} = sort(SelectedERs,$ 
        \Statex $C_{st}:ascendin g,ER_{i.Act\_Rwd}:descending)$ 
        \label{startph2}
        \State $compRel = 0;$ $compRwd = 0;$
        \State $Provider_{ST} = P_{st};$ $Provider_{EC} = P_{ec};$
        \For{$ER_{i}\; in\; SelectedERs_{sorted}$}
            \If{$C_{i.st}\geq Provider_{st}$}
                \State $compRel \mathrel{+}= C_{i.Rel}$
                \State $compRwd \mathrel{+}= ER_{i.Act\_Rwd}$
                \State $Provider_{st} = C_{i.et}$
                \State $Provider_{ec} \mathrel{-}= C_{i.re}$
                \State $ER_{comp}.insert(ER_{i})$
            \EndIf
        \EndFor
        \State \Return $ER_{comp}, compRel, compRwd$ \label{endph2} 
    \end{algorithmic}
\end{algorithm}

\subsection{Selection of Energy Service Requests}
\label{phase1algo}
The selection phase selects energy requests (ERs) that are composable with the provider’s energy service (EaaS). The selection phase has two stages: The first stage involves the spatio-temporal selection of energy requests (See Fig.\ref{ReliableEnergyFrameWork}). The second stage includes both the reliability score calculation and the rewards computation for the selected ERs. The first stage selects neighbor ERs that may be served by the provider. The selection is achieved using the spatial and temporal features of EaaS. The spatial classification is required to transfer the wireless energy between the consumers and the provider. The temporal selection of an ER requires for its duration to fall within the time window of the energy provider. The final step, in the first stage, is to check if the requested energy may be provided by the EaaS. The second stage computes the reliability score for each selected ER. We compute the reliability score using the reliability model discussed in Section \ref{RM}. We also use Eq.\ref{rewardEquation} to compute the reward in the second stage based on the amount of requested energy.

The selection phase of the energy requests is detailed in Phase 1 in Algorithm \ref{alg:RCER} (Lines \ref{startph1} - \ref{endph1}). For every ER, the algorithm checks if the time interval of the ER falls in the time window of the energy service (Lines 1 - 2). Then, the distance $d$ between the ER and the energy service is computed in Line 3. Line 4 checks if the computed distance $d$ is less than the maximum required distance to transfer energy (15 feet using Energous). Line 5 tests if the available energy service is enough to provide the requested energy, while Line 6 computes the reliability score of the consumer using the previously defined reliability model (Section \ref{RM}). Line 7 computes the reward value of the ER using Eq.\ref{rewardEquation}. Line 8 computes the actual reward using Eq.\ref{ActRwd}. Line 9 inserts the ER information, the consumer reliability score, and the ER's actual reward value in the set of selected ERs. 

\subsection{Reliability-based Composition of Energy Requests}
\label{phase2algo}

Phase 2 of algorithm \ref{alg:RCER} represents the composition of the energy  requests (Lines \ref{startph2} - \ref{endph2}). This phase aims to compose the most reliable ERs that ensure the maximization of the provider's reward. Line \ref{startph2} sorts the selected ERs in ascending order based on their start time, then in descending order based on their actual reward. Lines 13 - 15 check for each $ER$ if the available $EaaS$ of the provider can deliver the requested energy, then add up the consumer's reliability score, as part of the composition's reliability using Eq.\ref{compReliabilityEq}. In addition, Line 16 adds up the ER's actual reward $ER_{i.Act\_Rwd}$ as part of the provider's reward. Line 17 updates the start time of the provider so that ERs don't overlap, since the provider can only transfer energy to one consumer at a time. Line 18 updates the remaining energy of the provider by subtracting the amount of requested energy $C\_{re}$ from the provider's available energy $P\_{ec}$. Line 19 inserts the ER to the set of composed ERs.

\begin{algorithm}[t!]
\setlength{\textfloatsep}{-50pt}
\setlength{\floatsep}{0pt}
    \renewcommand{\algorithmicrequire}{\textbf{Input:}}
    \renewcommand{\algorithmicensure}{\textbf{Output:}}
    \caption{{Adaptive Reliability-Based Composition }}
    \label{alg:ARCER}
    \begin{algorithmic}[1]
        \Require
        $EaaS, All_{ER}, Reliability\; Model (RM) $
        \Ensure $ER_{composition} , compReliability , compReward$ 
        \Statex \textbf{Phase 1: Selection of Energy Service Requests}
        \For{$ER_{i}\; in\; All_{ER}$}\label{Astartph1} \do 
            \State \If{$ [C_{st},C_{et}] \subseteq [P_{st}, P_{et}]$}
          
                \State \small{$ d = Compute\_Distance(C_{loc.x},C_{loc.y},P_{loc.x},P_{loc.y}) $}
                         
                \If{$d \leq Max\;Energy\;Distance$ }
                    \If{$C_{i.re}\leq P_{ec}$}
                        \State{$C_{i.Rel.Voluntary} = Vol\_Rel(ER_{i},RM)$}
                        \State$C_{i.re} = C_{i.re} * C_{i.Rel.Voluntary}$
                        \State $C_{et} = C_{st} + (C_{et} - C_{st})* C_{i.Rel.Voluntary}$
                        \State $C_{i.Rel} = Compute\_Reliability(ER_{i},RM)$
                        \State $ER_{i.Reward} = C_{i.re} /P_{ec}$
                        \State $ER_{i.Act\_Reward} = C_{i.Rel} * ER_{i.Rwd}$
                         \State \small{$SelectedERs.insert(ER_{i} , C_{i.re},ER_{i.Act\_Rwd})$}\label{Aendph1}
                    \EndIf 
                \EndIf 
            \EndIf
        \EndFor
        \Statex \textbf{Phase 2: Composition of Energy Service Requests}
        \Statex \textbf{Similar to Phase 2 of  algorithm \ref{alg:RCER}}
        \State \Return $ER_{comp}, compReli, compRwd$ \label{Aendph2} 
    \end{algorithmic}
\end{algorithm}

\subsection{Adaptive Reliability-Based Composition of Energy Requests}

We propose an \textit{adaptive} reliability-based composition as an enhancement to Algorithm \ref{alg:RCER}. In the adaptive composition, ERs will be modified and divided based on their voluntary reliability. Recall from subsection \ref{RM}, voluntary reliability measures the commitment of users to fully receive their requested energy. Therefore, our approach will downsize the amount of ER based on the consumer's voluntary reliability score.  Algorithm \ref{alg:ARCER} describes the composition in detail, and is implemented across two phases: A selection phase and a composition phase. Lines \ref{Astartph1} - \ref{Aendph1} describe the first phase of selecting ERs that are composable with the provider's energy service. The selection phase has two stages: The first stage is similar to subsection \ref{phase1algo} (Lines 1 - 5 ). The second stage includes computing the reliability score, downsizing ERs, and computing the rewards for the selected ERs (Lines 6 - 11). Line 5 tests if the available energy service is enough to provide the requested energy. Line 6 computes the voluntary reliability score of the ER using Eq.\ref{voluntrayeq}. Line 7 downsize the mount of requested energy based on the voluntary reliability score. Similarly, Line 8 adjusts the ER's end time based on the voluntary reliability score. Line 9 computes the reliability score of the consumer using the previously defined reliability model (Subsection \ref{RM}). Line 10 computes the reward value of the ER using Eq.\ref{rewardEquation}. Line 11 computes the actual reward using Eq.\ref{ActRwd}. Line 12 inserts the ER's information, the consumer reliability score, and the ER's actual reward value in the set of selected ERs. The composition phase follows the same description of Phase 2 in subsection \ref{phase2algo}.

\begin{table}[!t]
\centering
\caption{Experiments Variables}
\label{ExpVariables}
\resizebox{\linewidth}{!}{%
\renewcommand{\arraystretch}{1}
\begin{tabular}{|l|l|}
\hline
Variables & Value   \\ \hline
Total Energy Requests for coffee shop 1 in April            & 16830   \\ 
Energy Services                  &{8000}             \\ 
Duration of Services             & {30 - 120 minutes} \\ 
Duration of Energy Requests      & {5 - 30 minutes}   \\
Provided Energy                  & {50 - 100 \%}      \\
Requested Energy                 & {1 - 100 \%}       \\
Battery Level                    & {1 - 60\%}         \\  \hline
\end{tabular}}%
\end{table}

\section{Experiments and Discussions}

\label{experiments}
We compare the proposed composition approaches, the Reliability-Based composition (RB), and the Adaptive Reliability-Based Composition (ARB), with two different approaches, a greedy approach (Greedy) and a Brute Force approach (BF). The greedy approach is inspired by the first come first served algorithm \cite{kruse2007data}, where the ERs will be sorted and composed based on their start time without considering their reliability. We consider the Greedy approach as the baseline.  In the Brute Force approach, we retrieve all the possible compositions of the energy service requests. Then, we select the composition that has the maximum total actual reward.

We evaluate the effectiveness and the computation cost of the composition approaches. The effectiveness of each composition approach is measured in terms of their reliability score, reward value, and energy utilization. The experiments only test the performance of the approaches from a \textit{single} provider perspective.

\subsection{Dataset Description}

 To the best of our knowledge, there is no existing dataset for wireless energy sharing among IoT devices. Therefore, we built an energy services environment close to reality, based on a dataset released by IBM for a coffee shop chain with three branches in New York city\footnote{https://ibm.co/2O7IvxJ}. The dataset consists of one-month transaction records of customers purchases in each coffee shop. Each coffee shop consists of, on average, 560 transnational records per day, and a total record of 16,500 transactions. We consider the coffee shop customers as IoT users in the prospected environment. We assume energy services are provided from wearables or from the spare energy of IoT users' batteries. Similarly, energy requests are sent by IoT users. Additionally, we assume that energy services and requests are sent to an IoT coordinator at an edge server. For example, a WiFi router in a coffee shop. In our experiments, we use the consumer ID, transaction date, time period, location, and coffee shop ID from each record in the dataset to define the spatio-temporal features of the energy services and requests. For example, the start and end time of energy service or a request. We augment the dataset by randomly generating the rest of the QoS parameters based on a uniform distribution.  For instance, the transactions' time and location are used as the energy requests starting time and location. The ending time of each energy request was generated randomly.  We used a random uniform distribution to generate the amount of provided energy for each service and the amount of requested energy for each request.  Moreover, we generated the energy disruptions for each request.  We ran a total of 8,000 energy service advertisements. For each energy service, we used our proposed approaches to compose energy requests. Table \ref{ExpVariables} summarizes the experimental variables. 
\begin{figure}[!t]
    \centering
   \includegraphics[width=0.85\linewidth]{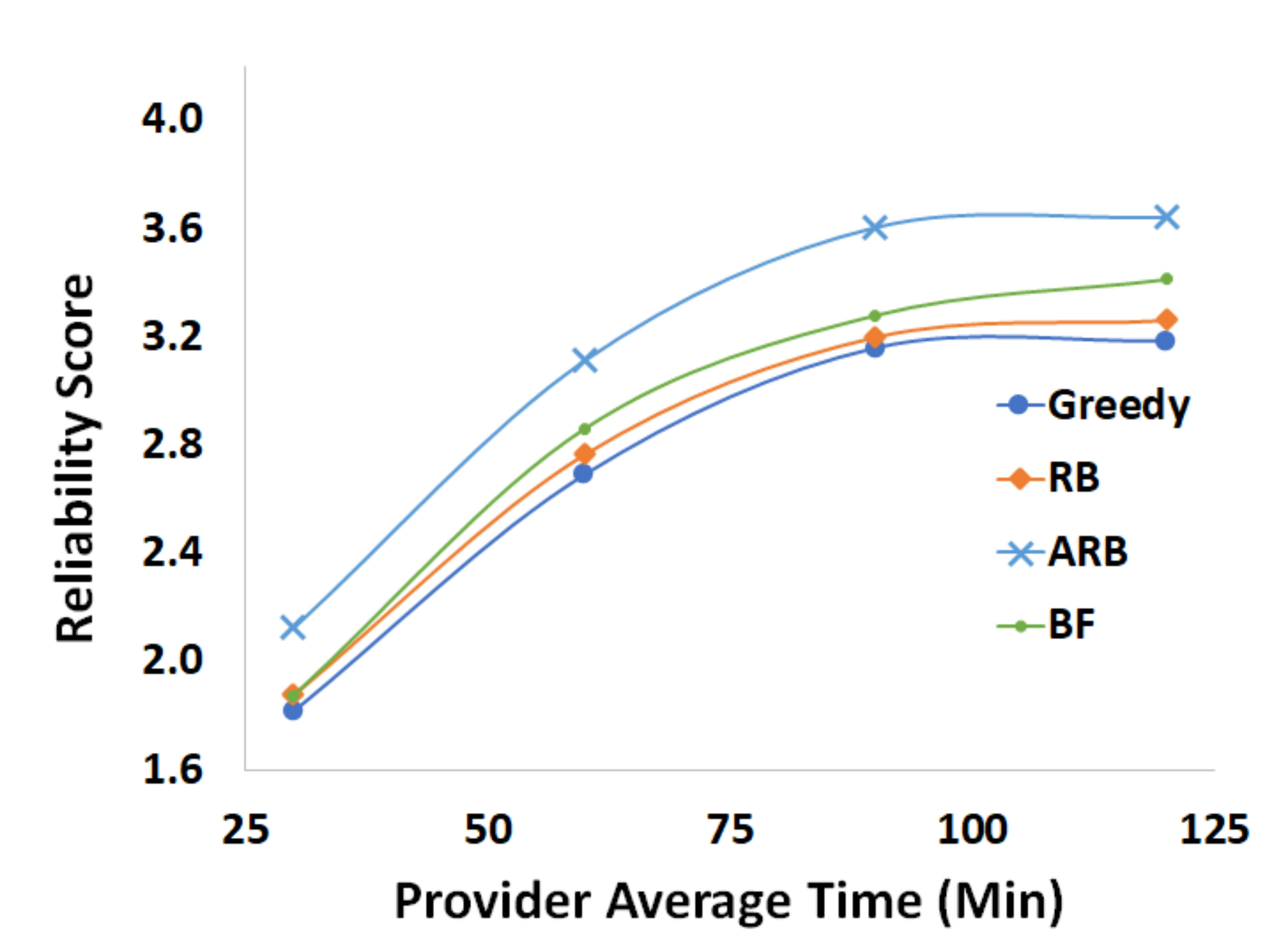}
    \caption{The average of the Reliability Score.}
    \label{ReliabilityFig}
\end{figure}
\subsection{Evaluation of the Composition Framework}
We ran four experiments to determine the efficiency of our proposed approaches. The experiments evaluated the composition approaches in terms of their reliability score, reward value, energy utilization, and computation cost. We run the composition algorithm and the other techniques in different settings by changing the time interval length of services.  We gradually increased the time duration of energy services from 30 minutes to 120 minutes. We repeated the experiment 2000 times at each point where the duration of energy services increased. We considered the average value for each technique at each point.

\begin{figure}[!t]
\centering
 \includegraphics[width=.85\linewidth]{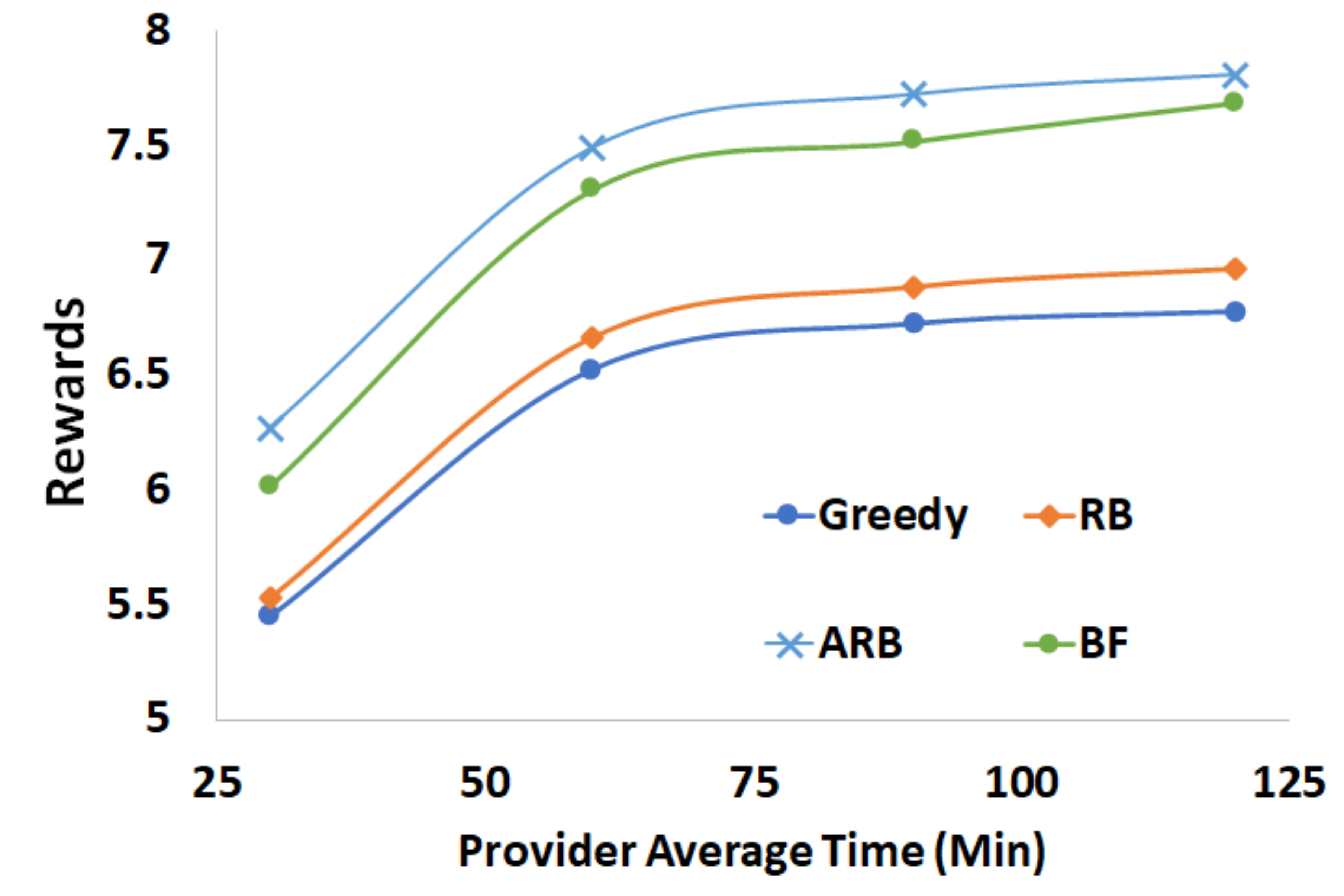}   
\caption{The average of total rewards.  }
\label{RewardAll}
\end{figure}

\begin{figure}[!t]
\centering
   \includegraphics[width=\linewidth]{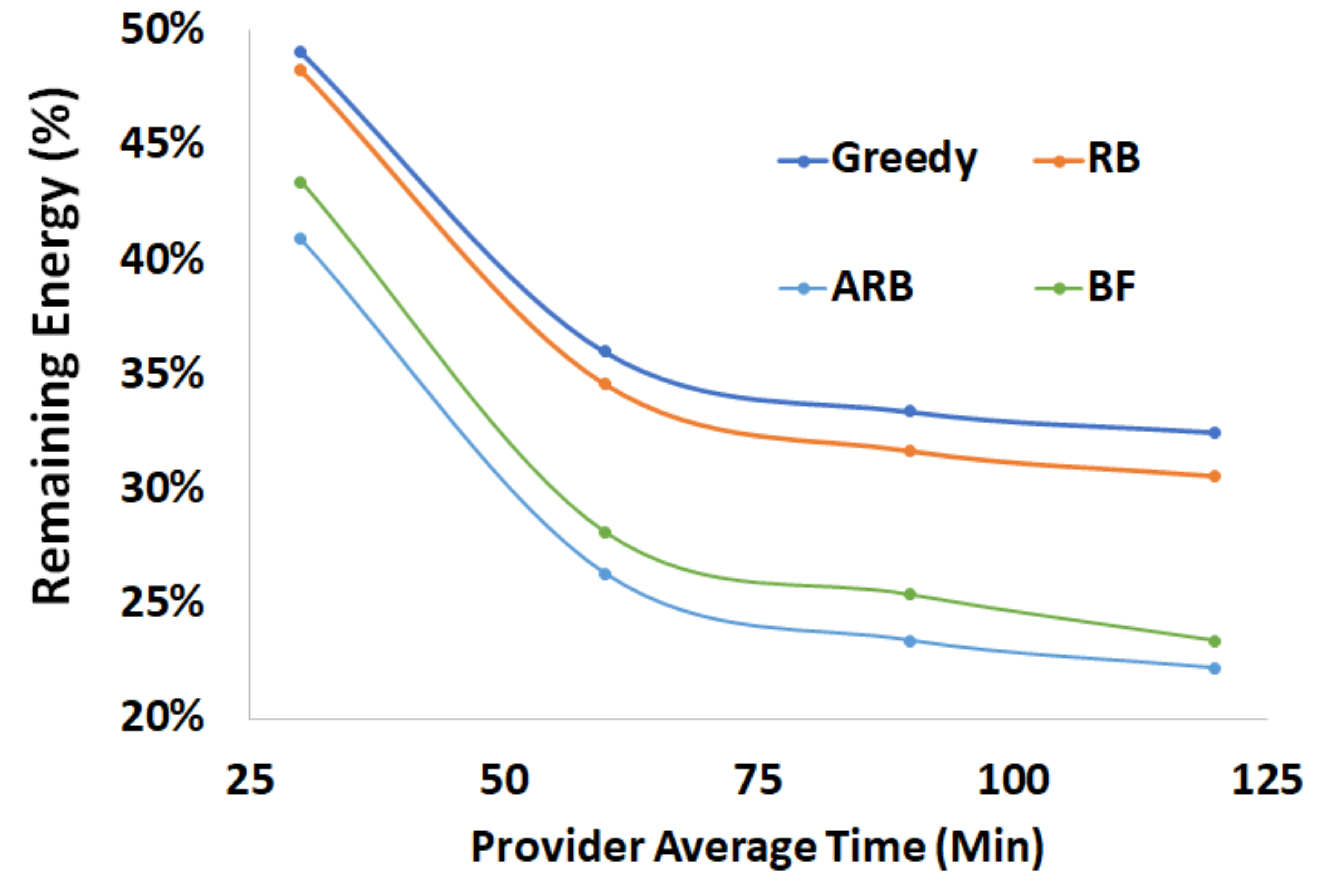}  
 \caption{The average of providers' remaining energy.}
    \label{RemainEnergy}
\end{figure}

\begin{figure}[!t]
\centering
 \includegraphics[width=\linewidth]{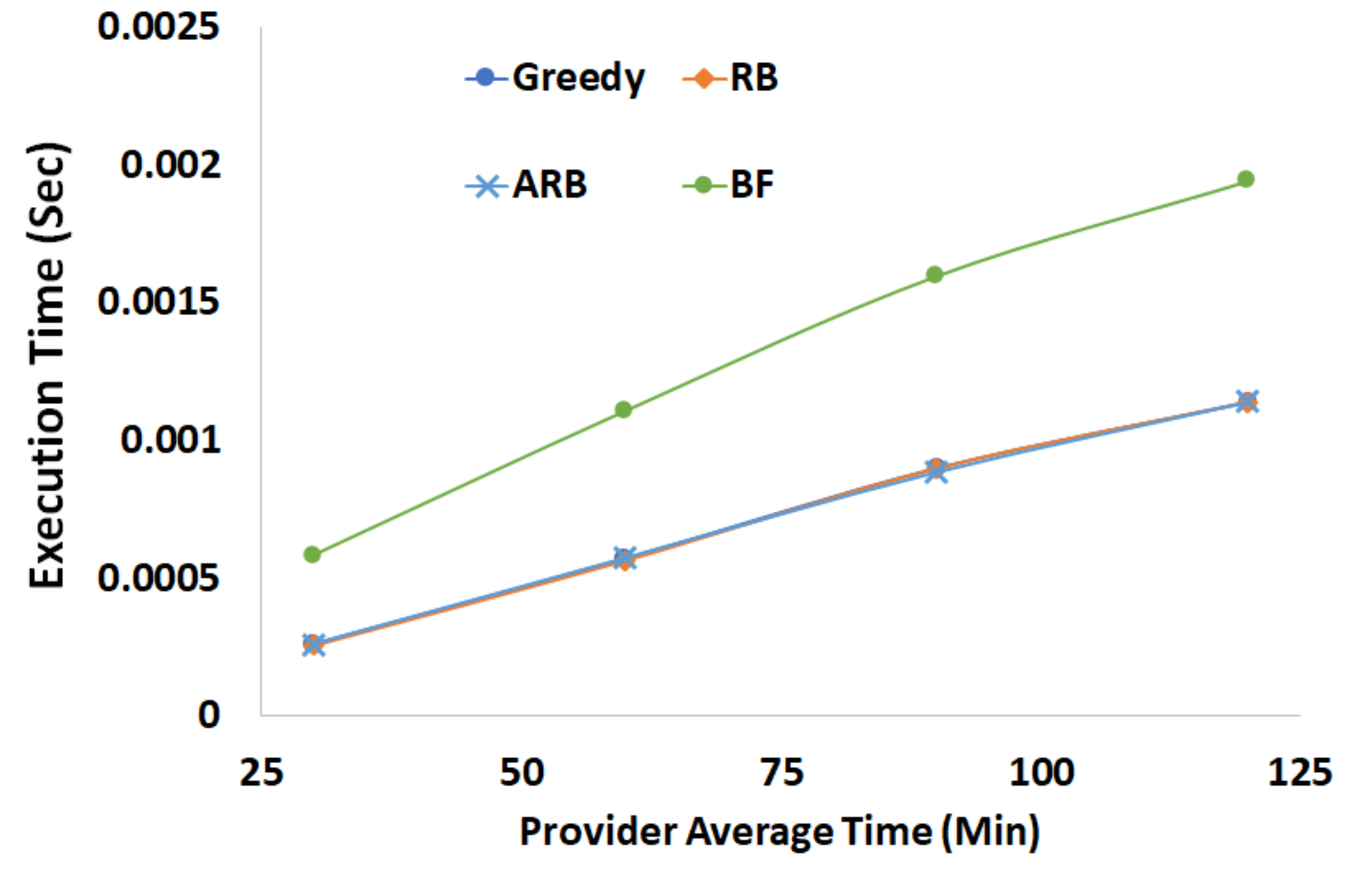}   
\caption{The average of execution time.  }
\label{ExecutionTime}
\end{figure}

The first experiment compares the average reliability score of the proposed composition algorithms RB and ARB, against BF and Greedy. As previously stated, the reliability score presents the commitment of the consumers to receive the requested energy. Therefore, a high-reliability score of a composition ensures a better reward for the provider. The composition reliability score is computed using  Eq.\ref{compReliabilityEq}. Fig.\ref{ReliabilityFig} presents the average reliability score for each algorithm. The x-axis in Figs.[\ref{ReliabilityFig}-\ref{RemainEnergy}] {represents the provider's average staying time}.  In Fig.\ref{ReliabilityFig}, the reliability score increases when the provider staying time increase for all the composition algorithms. For instance, when the provider staying time is 30 minutes, all algorithms provide a less reliability score compared to their scores when the provider staying time is 120 minutes. This observation can be explained by the provider staying time to share their energy. The longer that staying time of the provider, the more requests they can fulfill. Therefore, the aggregated reliability score of these requests will increase. The proposed algorithm RB performs slightly better than Greedy in terms of reliability as it considers the reliability of each ER, unlike the Greedy approach. In addition, the BF approach gives slightly better results than RB as it looks for all the possible combinations of compositions, and it selects the composition with the highest reliability score. However, the BF approach has a higher computation cost compared to RB as shown in Fig.\ref{ExecutionTime}. The proposed algorithm ARB gives the best results as it downsizes the amount of requested energy in each energy request based on their reliability score. Downsizing the requests allows the ARB algorithm to include more requests, and therefore aggregate additional reliability scores.

The second experiment compares the average of the total reward of each approach. Recall that rewards incentivize the provider to share their energy. Therefore, a high reward encourages providers to participate more in energy sharing. Similar to the previous experiment, in Fig.\ref{RewardAll}, the reward value increases when the provider's staying time increase for all the composition algorithms. For instance, when the provider staying time is 120 minutes, all algorithms provide more reward value compared to when the provider staying time is 30 minutes. Similarly, this observation can be justified by the provider's staying time to share their energy. The longer the staying time of the provider, the more requests they can fulfill, and therefore, the reward value increases. By observing Fig.\ref{RewardAll}, it is clear that the proposed approach RB results in higher rewards than the Greedy approach.  The BF approach gives better results compared to RB and Greedy. However, as aforementioned, BF requires more computations.  The proposed algorithm ARB gives the best results among all approaches. As previously mentioned, the ARB algorithm downsizes the amount of requested energy in each energy request based on their reliability score. Downsizing the requests allows the ARB algorithm to include more requests, and therefore aggregate additional rewards.

The third experiment compares the average of the providers' remaining energy after each composition (See Fig.\ref{RemainEnergy}).  A low remaining energy of a providers' energy service indicates a good energy utilization which promotes a green IoT environment. Therefore, the experiment aims to compare the energy utilization of each approach by comparing their remaining energy. In Fig.\ref{RemainEnergy}, the remaining energy decreases when the provider staying time increase for all the composition algorithms. For instance, when the provider staying time is 120 minutes, all algorithms provide a low remaining energy percentage compared to their scores when the provider staying time is 30 minutes. This observation can be explained by the provider's staying time to share their energy. The longer the staying time of the provider, the more requests they can fulfill. Thus, the provider's remaining energy will be less.  The figure also shows that the proposed approach RB performs better than the Greedy approach in terms of utilizing the provider's energy service. As aforementioned, the RB approach composes ERs while considering the reliability score of each ER.  Considering the reliability enables the RB approach to ensure the commitment of consumers to receive energy which results in better energy utilization. Comparing  BF to RB and Greedy, the BF method provides better results. However, BF needs further computations as previously mentioned. The proposed algorithm ARB gives the best results among all approaches. As previously mentioned, the ARB algorithm downsizes each energy request based on its reliability score.  Downsizing the requests allows the ARB algorithm to include more requests and therefore increases the utilization of providers' energy.  

The fourth experiment evaluates the computation cost of all the composition approaches. Fig.\ref{ExecutionTime} shows the execution time of each approach. The figure shows that the BF approach has the highest computation cost because the BF approach looks for all the possible combinations of compositions, and selects the composition with the highest reliability score. The ARB approach has a lower computation cost than the BF approach.  Additionally, The ARB approach offers the best results in terms of the reliability score, the total reward, and the energy utilization, as previously discussed and presented in Figs.[\ref{ReliabilityFig}-\ref{RemainEnergy}].

\section{Related work}

There is a growing interest in energy sharing among mobile devices \cite{dhungana2020peer}. Energy Sharing is used for several purposes, including utilizing energy, balancing energy distribution, and energy sharing for content delivery \cite{dhungana2020peer}. Energy utilization intents to charge from other devices' battery to reduce relying on outlets  \cite{dhungana2019exploiting}. Balancing energy distribution is accomplished in a mobile network by exchanging energy among devices \cite{nikoletseas2017wireless}, or by using a star network design \cite{madhja2018peer}. Additionally, energy sharing is used in content delivery where nodes forward messages to a destination node \cite{dhungana2019energy}. The message is transferred with energy to encourage the nodes to carry the content to the destination node.  Existing research presumes that mobile devices are motivated to part energy \cite{dhungana2020peer}. Incentives are needed to increase participation in energy provision. Using incentives while composing reliable energy services requests is yet to be addressed \cite{dhungana2020peer}.

Reliability of services has been widely explored in the literature including web, and mobile sensing services \cite{zheng2013personalized} \cite{an2015crowdsourcing} \cite{talasila2015mobile}. A proposed approach to determine the reliability of web services suggested two methods neighborhood-based and model-based \cite{zheng2013personalized}. The neighborhood-based approach uses past failure data of similar neighbor services to predict the Web service's reliability. The model-based approach uses a factor model built on the data of services failure, and then it uses this factor model to make a reliability prediction.  Another approach to evaluate the reliability of mobile sensing services proposes the use of a link reliability factor \cite{an2015crowdsourcing}. The factor determines the reliability of routes between a provider and a requester. Measuring the reliability in mobile sensing services, additionally, included the reliability of providers to produce accurate data \cite{varshney2012privacy}. Another work suggested measuring the reliability and accuracy of the provided service, by the repetition of performing a task and a majority voting system \cite{varshney2012privacy}. Another approach proposed a reliability assessment method to assign sensing tasks to vehicles \cite{halabi2019reliability}. The reliability method evaluates vehicles based on previously completed sensing tasks in terms of acceptable task performance and accuracy of sensed data. All the aforementioned approaches are not applicable to the EaaS environment because of the context of the EaaS environment, such as the mobility of IoT users and the resources' constrains. In addition, composing reliable energy services requests while using incentives is yet to be addressed \cite{dhungana2020peer}. This paper, hence, is the first attempt to select and compose reliable incentive-driven energy requests to increase the participation in provisioning energy.

Our proposed approach adds a new contribution to the existing literature through several aspects. First, the proposed framework composes requests while considering incentives to increase providers' participation in the energy provision. Moreover, the proposed reliability model studies the behavior of consumers in a dynamic crowdsourced IoT environment. In addition, the proposed Spatio-temporal composition framework uses the reliability model to select the optimal composition of energy requests. The composition of energy requests delivers the most reliable energy requests the maximize the provider's reward. These contributions are among the early attempts to compose crowdsourced energy requests in IoT environments \cite{dhungana2020peer}.

\section{Conclusion}

We proposed an energy requests' composition framework that evaluates consumers' reliability while incentivizing providers to share energy. A new reliability model was proposed to capture the users' consumption patterns. A reliability-based and an incentive-driven composition of energy service requests was proposed. The approach selects the most reliable energy service requests that maximize the reward of the provider to overcome the resistance of provision. We additionally proposed an adaptive approach where energy requests will be downsized based on the consumers' reliability score.  Experimental results show that the adaptive composition (ARB) outperforms all the evaluated approaches. The efficiency of the proposed approaches was tested against the Brute  Force (BF) and the Greedy approaches. Future direction is to accommodate different incentive preferences in the framework and compose requests for multiple providers. 
\bibliographystyle{IEEEtran}
\bibliography{ref}

\end{document}